\documentstyle[aps,pre]{revtex} 
\input epsf
\begin{document}                       
\draft
\twocolumn

\title{Kinetics of Modulated and Ordered Structures in CuAu.}

\author{K. R. Elder$^{1,4}$, Nicholas A. Gross$^2,3$, Bulbul Chakraborty$^3$ 
and Nigel Goldenfeld$^4$.}

\address{$^1$ Department of Physics, Oakland University, Rochester, MI,
48309-4401}
\address{$^2$ Boston University, College of General Studies,
981 Commonwealth Ave., Boston, MA 02215.}
\address{$^3$ Physics Department, Brandeis University, Waltham, MA
02254.} 
\address{$^4$ Department of Physics,
University of Illinois at Urbana-Champaign,
1110 West Green Street, Urbana, Illinois 61801.} 

\date{\today}

\maketitle

\begin{abstract}

A continuum model derived from an atomistic Hamiltonian
is used to examine the ordering kinetics in CuAu.  
A detailed description of the formation of the low and 
high temperature ordered and modulated superlattice states 
is given.  The metastability of the modulated phase 
at low temperatures is shown to severely hinder creation 
of the ordered superlattice. Formation of the modulated 
superlattice is shown to results in interesting lamellar 
and labyrinthine structures.

\end{abstract}

	The vast majority of synthetic and naturally occurring materials
contain long-lived non-equilibrium morphologies which
determine material properties. Understanding  
the non-equilibrium or kinetic processes that lead to such 
microscopic structures is, therefore, one of the most important tasks facing
materials theory.  The binary alloy CuAu is an excellent material 
to study such processes for the following reasons: the existence of 
a modulated superlattice at intermediate temperatures
gives rise to a rich set of phase transformations and 
microstructures; the kinetics are accessible in x-ray scattering 
experiments; and a continuum model for the kinetics has been derived 
from a quantum mechanical description of this alloy \cite{cx92}.    
In this paper the continuum model will be used to provide a 
detailed theoretical description of the non-equilibrium processes 
that occur in CuAu.  Thus predictions that can be verified 
experimentally are obtained from a microscopic description of 
the alloy.

The continuum model predicts two first 
order phase transitions; one from a disordered superlattice to 
a modulated superlattice at $T=T_{MD}$ and the other from the modulated 
phase to an ordered superlattice at $T=T_{OM}$ 
(where $T_{OM} < T_{MD}$).
The purpose of this paper is to determine the transient morphologies 
that arise following instantaneous temperature quenches that bring 
the system from one equilibrium state to another.
The results of this study indicate that: the ordered 
superlattice is difficult to form just below $T_{OM}$ 
due to the creation of a long-lived metastable modulated phase; 
the growth of an ordered phase far below $T_{OM}$ is hindered 
by the presence of `terminal' droplets; and the nucleation of the 
modulated phase from the low and high temperature phases leads to 
labyrinthine and lamellar patterns respectively.  

The appearance of the modulated superlattice was predicted 
from an atomistic model by Chakraborty and Xi\cite{cx92}.  In this work,
effective medium theory\cite{jnp87,j88} was employed to obtain an 
approximate classical Hamiltonian from a full quantum mechanical model. 
A mean field analysis of this Hamiltonian was used to obtain 
the free energy as a function of the average sublattice 
concentration ($\eta$) along one degenerate ordering direction. 
This degeneracy can lead to interesting effects
but will not be considered here.  The free energy can 
be written,
\begin{eqnarray}
\label{eq:free}
 {\cal F}(\eta) = {\cal F}_o &+& (2\pi/a)^d\int d\vec{r} [a_T \eta^2
-u\eta^4+v\eta^6 \nonumber \\
&-&e|\vec{\nabla}_\perp \eta|^2+e|\vec{\nabla_{||}}|^2
+f(\nabla_\perp^2\eta)^2 ],
\end{eqnarray}
where, $a_T=0.042\Delta T meV/K$, $\Delta T = T-T_o$, $T_o=1219 K$, 
$u=5.25meV$, $v=6.1meV$, $e=25.5meV/(2\pi/a)^2$ 
and $f=195.5meV/(2\pi/a)^4$ and $a$ is the shortest lattice parameter 
in the tetragonal structure of the low temperature CuAu(I) ordered
phase.  The symbols $\nabla_\perp$ and 
$\nabla_{||}$ refer to gradients in the plane 
perpendicular and parallel to the ordering direction respectively.
At low, intermediate and high temperatures 
${\cal F}$ is minimized by ordered (i.e., $\eta(\vec{x}) = \eta_o \neq 0$),
modulated (e.g., $\eta(x,y,z)=\eta(x+\lambda,y,z)$)
and disordered (i.e., $\eta(\vec{x})=0$) states respectively.
The transitions from the ordered to modulated phase and 
from the modulated to disordered phase are both 
first order and respectively occur at $\Delta T_{OM} \approx -8.0K$, 
and $\Delta T_{MD} \approx 44.4K$\cite{cx92,ceg95}.
The spinodal temperatures $T_O^S$ and $T_M^S$ are defined as 
the temperatures above which the ordered and 
modulated phases are unstable, while $T_D^S$ is defined as 
the temperature below which the disordered state is unstable to the modulated
phase.  
The values of these temperatures are $T_O^S=T_o+35.7K$, 
$T_M^S=T_o+52.6K$ and $T_D^S=T_o+20.0K$.

The kinetics are assumed to be relaxational and driven by minimization
of the free energy: 
\begin{equation}
\label{eq:eom}
\partial \eta / \partial t = -\Gamma \delta {\cal F} /\delta \eta + \zeta,
\end{equation}
where $\zeta$ is a random Gaussian noise term with correlations,
$\left<\zeta(\vec{r},t)\zeta(\vec{r}',t')\right> 
= 2k_B T \Gamma \delta(\vec{r}-\vec{r}')
\delta(\tau-\tau')$ and $k_B$ is the Boltzmann constant.  It is
convenient to introduce the scaled variables;
$\vec{x}=\vec{r}/\lambda_o$ and $\tau=t/t_o$ where $\lambda_o = \sqrt{2f/e}$
and $t_o = (a/2\pi)^d2f/(e^2\Gamma)$ to obtain;
\begin{eqnarray}
\label{eq:eomd}
\partial \eta / \partial \tau = -(\gamma \Delta T 
+ 2\nabla_\perp^2&+&\nabla_\perp^4-2\nabla_{||}^2)\eta \nonumber \\
&+&u'\eta^3-v'\eta^5+\nu.
\end{eqnarray}
Here, $\gamma=4 a_T f/e^2$, $u'=8fu/e^2$, $v'=12fv/e^2$,
$\left<\nu(\vec{x},\tau)\nu(\vec{x}',\tau')\right> 
= 2\epsilon \delta(\vec{x}-\vec{x}')
\delta(\tau-\tau')$ and $\epsilon=2k_BT[a/2\pi]^d[2f/e^2][e/2f]^{d/2}$.  
In these rescaled units the modulated wavelength is simply $2\pi$.  
To simplify calculations ordering in the parallel direction 
will be neglected.  

	The kinetics that follow a rapid quench are influenced
by the modulated phase, even when the pre- and post-quench temperatures 
are not within the modulated regime.  For example, consider the 
instantaneous quench 
of a disordered state ($T_{initial} > T_{MD}$) 
into the ordered regime 
($T_{final} < T_{OM}$).  The initial response will be to 
form a modulated type structure, since linear stability analysis predicts
that the mode with $q= 1$ has the largest growth rate\cite{bc94}.
For $T_{final} < T_{OM}$ the growth 
of the $q=1$ mode is much more rapid 
than $q=0$, thus a `modulated' structure is quickly formed.  
Further growth of the $q=0$ mode (i.e., the ordered phase) is inhibited 
by the metastability of the modulated phase.  Thus for quenches just 
below $T_{OM}$ a long-lived transient modulated phase should develop.

\begin{figure}[btp]
\epsfxsize=3.4in \epsfysize=3.4in
\epsfbox{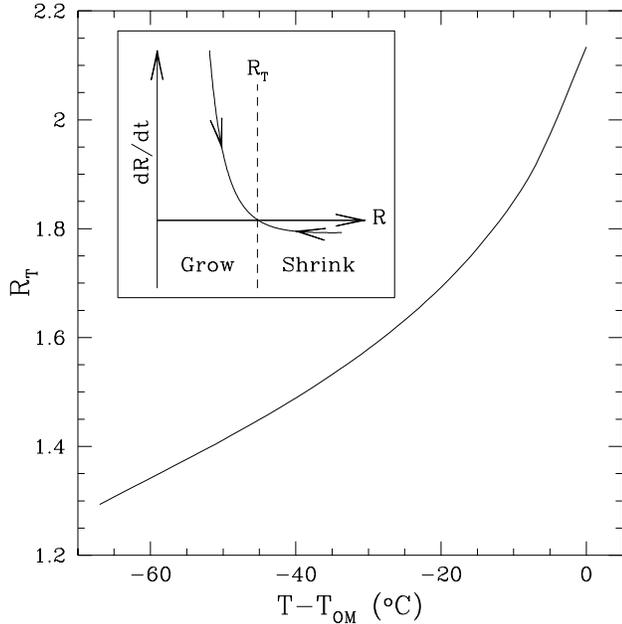}
\caption{In this figure the terminal droplet size is shown as
a function of temperature.  The inset
depicts the velocity of the droplet radius as a function of $R$ as
described by Eq. (4).}
\end{figure}

For $T_{final} \ll T_{OM}$ the difference between the $q=1$ and $q=0$ 
linear growth rates is insufficient to create a modulated 
structure.  The subsequent dynamics do not however reduce to 
that observed in standard order/disorder transitions in 
which the average ordered domain size grows as 
$t^{1/2}$\cite{ac75,gms83} and droplets shrink at a rate 
proportional to the curvature.  Here a droplet 
refers to a spherical (or circular in 2-d) regime of 
one phase (e.g., $\eta=-\eta_o$) embedded in 
the other (e.g., $\eta=+\eta_o$). 
In CuAu, large droplets shrink but do not disappear.
This effect can be traced to the fourth order 
spatial derivative in the dynamics or equivalently, to the existence of the
length scale describing the modulated phase, 
and to the two dimensional 
nature of the ordering process.

	To illustrate this effect, consider the dynamics of a single droplet 
of radius $R$ in the limit $R \gg W$, where $W$ is the domain wall thickness 
and is of the order $1$.  In this limit $\eta$ can be written 
$\eta(r-R(\tau)) \approx \eta^{1d}(r-R(\tau))$, where $\eta^{1d}(x)$ 
is the one-dimensional solution of $\delta {\cal F}/\delta \eta = 0$ 
with boundary conditions $\eta(x=\pm \infty)=\pm \eta_o$. 
Substituting this result into Eq. (\ref{eq:eomd}),
multiplying by $(\partial \eta^1d/\partial r)$ and integrating from 
$r=0$ to $r=\infty$ gives, 
\begin{equation}
\label{eq:drdt}
\partial R/\partial \tau = 2(1+\sigma_3/\sigma_1)R + 1/R^3,
\end{equation}
where $\sigma_i \equiv \int_0^{\infty} dr 
(\partial \eta^1d/\partial r) (\partial^i \eta^1d/\partial r^i)$.
The quantity $2(1+\sigma_3/\sigma_1)$ is less than zero 
for $T < T_{OM}$.  Thus 
the droplet will shrink for $ R > R_T$ and grow 
for $R < R_T $, where $R_T$ is the `terminal' droplet size 
and is equal to $R_T = 1/\sqrt{-2(1+\sigma_3/\sigma_1)}$. 
The growth rate of $R$ is illustrated in the inset of Fig. (1).
While this calculation correctly identifies the existence 
of a terminal droplet, the accuracy is hindered by the approximation 
$R\gg W$.  To overcome this obstacle $R_T$ was determined 
numerically and the results are shown in Fig. (1).  

\begin{figure}[btp]
\epsfxsize=3.4in \epsfysize=3.4in
\epsfbox{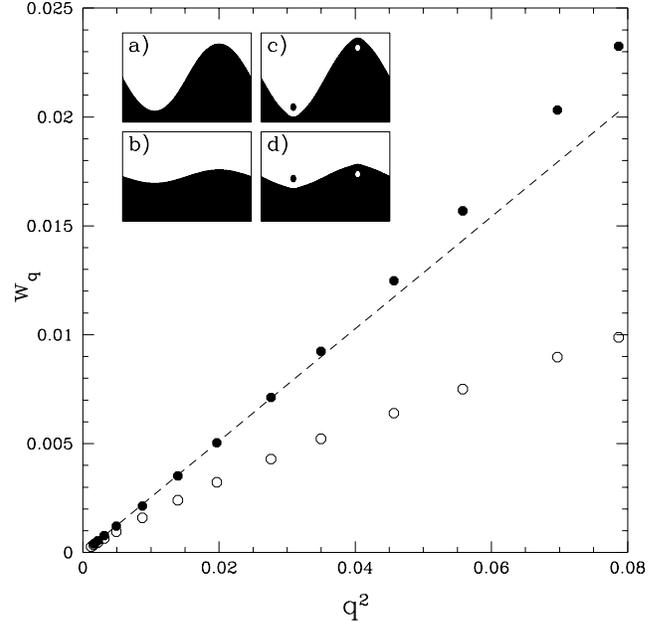}
\caption{In this figure the linear dispersion for the position
of an antiphase domain boundary is shown as a function of wavevector.
The solid and open points represent $w_q$ in the absence and
presence of droplets respectively and the dashed line corresponds to 
$w_q \sim q^2$.  In the inset figures (a) and (b) show the
interface relaxation in the absence of droplets and
in (c) and (d) in the presence of droplets at the same times.}
\end{figure}

	The terminal droplets influence the late stage dynamics 
by interfering with the motion of antiphase domain walls or interfaces.
For comparison it is useful to first consider the interface motion 
in the absence of droplets.  Consider a slowly varying interface 
such that $\eta(x,y,\tau) \approx \eta_o$ for $ x > h(y,\tau)$ and 
$\eta(x,y,\tau) \approx -\eta_o$ for $x < h(y,t)$, thus defining the 
position of the antiphase domain wall as $h(y,t)$.  If $h(y,t)$ varies 
slowly in space it is simple to show\cite{ceg95} that  
$\partial h /\partial \tau = [-2(1+\sigma_3/\sigma_1)\nabla^2 
+\nabla^4]h$ using the methods discussed in the previous paragraph. 
The solution in Fourier space is then;  
$\hat{h}(q,\tau) = e^{-w_q \tau}\hat{h}(q,0)$, 
where $w_q = -2(1+\sigma_3/\sigma_1)q^2+q^4$.  
In the long wavelength limit ($q\ll 1$) this reduces to the standard 
result for order/disorder transitions (i.e., $w_q \sim q^2$).
To determine the dispersion relationship (i.e., $w_q$) in the 
presense of terminal droplets numerical simulations were conducted 
for the configurations shown in the inset of Fig. (2). 
The results of these calculations (see Fig. (2)) indicate 
that the relaxation of the antiphase domain walls is restrained 
by the terminal droplets.  This 
simple effect will strongly alter the late stage dynamics as 
the droplets tend to accumulate at interfaces\cite{ceg95}.  
This accumulation is not due to an attraction between drops and 
interfaces,  but rather by droplets getting swept up by a relaxing 
interface.

	The transient morphologies that emerge in the modulated 
regime depend on both the pre- and post-quench states.  If the 
pre-quench temperature is above $T_{MD}$ and the post-quench temperature 
is between $T_{OM}$ and $T_D^S$, $\eta$ is linearly unstable and a convoluted 
modulated structure will quickly emerge.  Similar behavior will 
be observed if the pre-quench temperature is below $T_{OM}$ and the 
post-quench temperature is between $T_O^S$ and $T_{MD}$.  
In contrast, the kinetics are dominated by nucleation and growth 
when the pre-quench state is metastable at the post-quench 
temperature.  In addition growth of the nucleated 
droplets is not a simple process since there is an internal
structure within the drops.  Each drop will contain a lamellar 
structure that defines the modulated phase. In general the velocity 
of the droplet fronts will depend on the orientation of lamella 
with respect to the droplet front and on the background matrix 
the droplet is growing in.  To understand this effect it is useful 
to estimate the growth velocities in directions perpendicular and 
parallel to the lamella as the drop grows into either disordered 
or ordered backgrounds.

	First consider the invasion of the modulated phase into 
the ordered or disordered phase in a direction parallel to the 
lamella. The velocity of this front will 
be denoted $v^O_{||}$ and $v^D_{||}$ for propagation into the 
ordered and disordered phases respectively. To estimate 
$v^D_{||}$, $\eta$ can be approximated as 
$[\eta_o+\eta^{1d}(y)]F^O_{||}(x,\tau) - \eta_o$,
where $\eta_o$ is the value of $\eta$ in the ordered phase, 
$\eta^{1d}(y)$ is the one dimensional modulated solution 
(i.e., $\eta^{1d}(y) \approx A\sin(q_oy)$) and the overlap function 
$F^D_{||}(x,\tau)$ takes the values $1$ and $0$ in the modulated and 
ordered phases respectively.  Substituting this approximation 
into Eq. (\ref{eq:eomd}), multiplying by $\eta^{1d}(y)$, 
and averaging over one wavelength in $y$ gives, 
\begin{eqnarray}
\label{eq:fO||}
\partial F^O_{||} /\partial \tau = &(&1-\gamma \Delta T 
+3\eta_o^3u'-5\eta_o^4v'- \partial_{xxxx})F^O_{||} \nonumber \\
&+&2\eta_o^2(10v'\eta_o^2-3u')(F^O_{||})^2 \nonumber \\
&+&3(\eta_o^2+A^2/4)(u'-10\eta_o^2v')(F^O_{||})^3 \nonumber \\
&+&5v'\eta_o^2(4\eta_o^2+3u'A^2) (F^O_{||})^4  \nonumber \\
&-&5v'(\eta_o^4+3\eta_oA^2/2+A^4/8)(F^O_{||})^5. 
\end{eqnarray}
if $\eta^{1d}$ is approximated by a single Fourier mode.
The front velocity can be obtained by expanding around the zero 
velocity limit which occurs at $T=T_{OM}$. Substituting the 
approximation, 
$F^O_{||}(x,t)\approx f^O_{ll}(x-v^O_{||}t)$ (were $f^O_{||}$ is 
the solution of Eq. (\ref{eq:fO||}) at $T=T_{OM}$, with boundary 
conditions $f^O_{||}(-\infty) = 1$ and $f^O_{||}(\infty)=0$)) 
into Eq. (\ref{eq:fO||})
and integrating over $\partial f^O_{||}(x)/\partial x$ gives,
\begin{equation}
v^O_{||} = 0.045(T-T_{OM})/\sigma^O_{||},
\end{equation}
where $\sigma^O_{||} \equiv \int dx (\partial f^O_{||}(x)/\partial x)^2$.
The same techniques can be used to estimate $v^D_{||}$ and leads to 
$v^D_{||} = 0.025(T_{MD}-T)/\sigma^D_{||}$, where 
$\sigma^D_{||} \equiv \int dx (\partial f^D_{||}(x)/\partial x)^2$,
and $f^D_{||}(x)$ is the solution of Eq. (\ref{eq:fO||}) with 
$\eta_o=0$ at $T=T_{MD}$ and boundary conditions 
$f^D_{||}(-\infty) = 1$ and $f^D_{||}(\infty)=0$.
Numerically determining $\sigma^O_{||}$ and $\sigma^D_{||}$ gives,
$v^O_{||} \approx 0.13 (T-T_{OM})$ and $v^D_{||} \approx 0.12 (T_{MD}-T)$.

	Next consider growth in the direction perpendicular to the 
lamella.  For these calculations the 
velocities will be denoted $v^O_{\perp}$ and $v^D_{\perp}$ for 
growth into the ordered and disordered regimes respectively.
To estimate $v^D_{\perp}$, $\eta$ can be approximated as;
$\eta(x,\tau)\approx \eta^{1d}(x)F^D_{\perp}(x,\tau)$.  Substituting 
this expression into Eq. (\ref{eq:eomd}), multiplying by 
$\eta^{1d}(x)$, averaging over one wavelength and assuming
that $F^D_{\perp}(x,\tau)$ varies slowly in space relative to 
the modulated wavelength gives,
\begin{eqnarray}
\partial F^D_{\perp} /\partial \tau = &(&1-\gamma\Delta T + 4\partial_{xx}
-\partial_{xxxx})F^D_{\perp} \nonumber \\
&+&(3u'A^2/4)(F^D_{\perp})^3 -(5v'A^4/8)(F^D_{\perp})^5
\end{eqnarray}
if a one mode approximation for $\eta^{1d}$ is used.
Employing the method outlined above gives,
$v^D_{\perp} = 0.025(T_{MD}-T)/\sigma^D_{\perp}$.  Determining 
$\sigma^D_{\perp}$ numerically gives $v^D_{\perp} \approx 0.19(T_{MD}-T)$.
It is much more difficult to estimate $v^O_{\perp}$ since 
the solution can not be written as an overlap function times the 
modulated solution.  Numerical 
attempts to determine $v^O_{\perp}$ indicate this velocity is extremely
small, in fact no growth in this direction was observed.  Thus to 
the accuracy of the current work
$v^O_{\perp} \approx 0$.  The reason for this slow growth is that
the modulated solution is very similar to the amplitude of the ordered 
solution and consequently the spatial transition from the 
ordered to modulated phase can be extremely sharp.  In the 
limit of an infinitely sharp interface $\sigma \rightarrow \infty$ and 
$v \rightarrow 0$.

	The results of these calculations indicate that for quenches 
from the disordered phase the droplets 
will grow asymmetrically in a direction perpendicular to the lamella,
leading to elongated lamella structures since 
$v^D_{||} < v^D_\perp$.
In contrast, for quenches 
from the ordered phase $v^D_{||} > v^D_\perp$.  
Thus a small spherical  
nucleated droplet will tend to sprout arms which will invade the 
neighboring territory.  This growth leads to a labyrinth type structure 
similar to that observed in reacting chemical fronts\cite{pg94}. 

	To illustrate the remarks of the preceding paragraphs 
several numerical simulations were conducted, which are illustrated 
in Fig. (3).  The numerical algorithm is discussed in a previous 
paper\cite{ceg95} and is based on the cell dynamics method\cite{op87}.  
Figures (3a) and (3b) respectively show transient patterns 
that emerge from quenches from the disordered regime to just below 
and far below $T_{OM}$.  The highly interconnected morphology
shown in Fig. (3a) is a long-lived metastable modulated structure that 
typically evolves only near defects.  The deeper quench (Fig. (3b))
shows that regions of high interface curvature coincide 
with large concentrations of terminal droplets, thus indicating 
the droplets inhibit interface relaxation and domain growth.
Figures (3c) and (3d) are transient patterns obtained by the nucleation
of the modulated phase from quenches from the disordered and ordered regimes
respectively.  These figures show the lamellar and labyrinthine type 
structures predicted in the preceding paragraphs.

\begin{figure}[btp]
\epsfxsize=3.4in \epsfysize=3.4in
\epsfbox{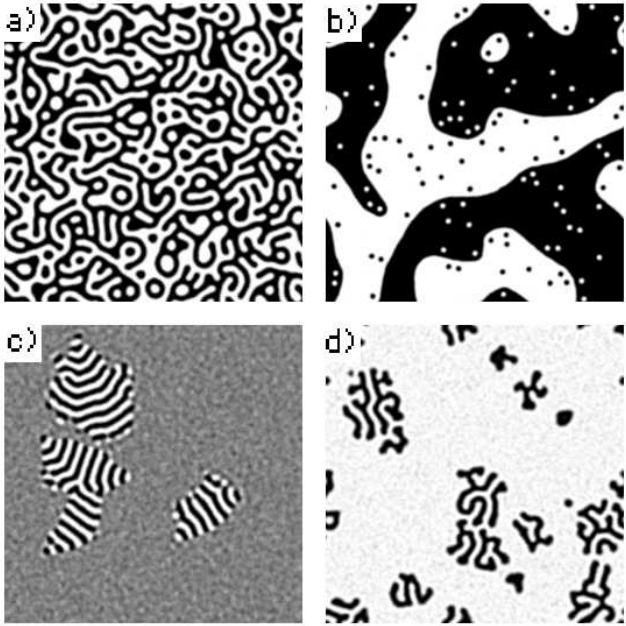}
\caption{Figures (a) and (b)
are the transient patterns that emerge following quenches
from a disordered state.  Figures (a) and (b) are at $\tau=182$ and
$\tau=329$ at quench temperatures of $T =T_{OM} -4K$ and
$T = T_{OM}-44K$ respectively.
Figure (c) is a pattern obtained
by quenching from the disordered state into the modulated regime
at $T=T_{OM}+39.5K$ at $\tau=57.3$.
Figure (d) is a pattern obtained
by quenching from the ordered state into the modulated regime
at $T=T_{OM}+38K$ at $\tau=84.0$.}
\end{figure}

In summary several predictions for the kinetics of phase transformations 
in CuAu have been made.  For quenches from the disordered to just 
below the ordered/modulated transition it is predicted that linear 
dispersion will produce a modulated morphology.  This modulated structure
will be long-lived since the modulated phase is metastable in the
ordered regime. For quenches far below $T_{OM}$ it was found that 
the formation of the ordered phases are strongly inhibited by the 
appearance of terminal drops.   Growth in this region will be much 
slower than standard order/disorder transitions since the droplets 
collect at antiphase boundaries and inhibit motion. 
For quenches into the modulated phases it was found the initial state 
played an important role in determining the transient 
morphologies.  Quenches from the ordered and disordered region 
respectively produced labyrinthine and lamellar type structures.  

	Finally it is interesting 
to speculate on the asymptotic dynamics in the modulated regime. 
In this case the dynamics are controlled by the elimination of defects 
and the relaxation of the orientation of the individual 
lamella.  Similar dynamics occur in the Swift Hohenberg\cite{sh77} model 
of Rayleigh-B\'enard convection 
and lead to the growth of domains of the same orientation growing at 
at rate of $t^{1/4}$ in the presence of thermal fluctuations and 
$t^{1/5}$ in the absence of fluctuations\cite{evg93}.  It is 
likely the same growth exponent would be measured in this system. 

The authors would like to thank Bill Klein, Oana Malis and 
Karl Ludwig for many useful discussions.  KRE 
acknowledges support of grant NSF-DMR-8920538
administered through the U. of Illinois Materials Research
Laboratory.  The work of BC and NAG was supported by NSF grants  
DMR-9208084 (BC) and DMR-952093 (NAG).  NG acknowledges 
support from NSF though grant DMR-93-14938.

\end{document}